\documentstyle[12pt,epsf,rotating]{article}
\newcommand{\lsim}{\raisebox{-0.13cm}{~\shortstack{$<$ \\[-0.07cm] $\sim$}}~}
\newcommand{\gsim}{\raisebox{-0.13cm}{~\shortstack{$>$ \\[-0.07cm] $\sim$}}~}

\newcommand{\ra}{\rightarrow}
\newcommand{\ee}{e^+e^-}
\newcommand{\s}{\\ \vspace*{-2.5mm} }
\newcommand{\nn}{\noindent}
\newcommand{\non}{\nonumber}
\newcommand{\beq}{\begin{eqnarray}}
\newcommand{\eeq}{\end{eqnarray}}

\newcommand{\MSSM}{MSSM}

\newcommand{\tb}{\mbox{tg}\beta}

\newcommand{\GeV}{GeV}

\begin{document}

\begin{titlepage}

\begin{flushright}
DESY 97-079\\
IFT-96-29\\
PM--97/04 \\
April 1997 
\end{flushright}

\def\thefootnote{\fnsymbol{footnote}}

\vspace{1cm}

\begin{center}

{\Large\sc {\bf  HDECAY:
}}

\vspace*{3mm}

{\large\sc {\bf  a Program for Higgs Boson Decays}}

\vspace*{3mm}

{\large\sc {\bf  in the Standard Model and its Supersymmetric Extension}}

\vspace{1cm}

{\sc A.~Djouadi$^{1}$, J. Kalinowski$^{2}$ and M. Spira$^3$ }

\vspace{.5cm}

$^1$ Laboratoire de Physique Math\'ematique et Th\'eorique, UPRES--A 5032, \\
\vspace*{1mm}
Universit\'e de Montpellier II, F--34095 Montpellier Cedex 5, France. \\
\vspace*{3mm}
$^2$ Deutsches Elektronen--Synchrotron, DESY, D--22603 Hamburg,
Germany, \\ 
\vspace*{1mm} Institute of Theoretical Physics,  
Warsaw University, PL--00681 Warsaw, Poland. \\
\vspace{0.3cm}
$^3$ Theory Division, CERN, CH--1211, Geneva 23, Switzerland.\\

\end{center}

\vspace{.9cm}

\begin{abstract}
\normalsize

\nn We describe the Fortran code HDECAY\footnote{The program may be
obtained from http://wwwcn.cern.ch/$^\sim$mspira/ or
http://www.lpm.univ-montp2.fr/$^\sim$djouadi/program.html, or via
E-mail from: djouadi@lpm.univ-montp2.fr, kalino@desy.de,
spira@cern.ch.}, which calculates the decay widths and the branching
ratios of the Standard Model Higgs boson, and of the neutral and
charged Higgs particles of the Minimal Supersymmetric extension of the
Standard Model. The program is self-contained (with all subroutines 
included), easy to run, fast and calculates the decay widths and 
branching ratios according to the current theoretical knowledge. 

\end{abstract}

\end{titlepage}

\def\thefootnote{\arabic{footnote}}
\setcounter{footnote}{0}
\setcounter{page}{2}

\section{Introduction}

The experimental observation of scalar Higgs particles is crucial for 
our present understanding of the mechanism of electroweak symmetry breaking.
Thus the search for Higgs bosons is one of the main entries in the
LEP2 agenda, and will be one of the major goals of future colliders
such as the Large Hadron Collider LHC and the future Linear $\ee$ Collider
LC. Once the Higgs boson is found, it will be of utmost importance to
perform a detailed investigation of its fundamental properties, a crucial
requirement to establish the Higgs mechanism as the basic way to
generate the masses of the known particles. To this end, a very precise
prediction of the production cross sections and of the branching ratios
for the main decay channels is mandatory. \s

In the Standard Model (SM), one doublet of scalar fields is needed for
the electroweak symmetry breaking, leading to the existence of one
neutral scalar particle $H^0$ \cite{HHG}. Once $M_{H^0}$ is fixed, the
profile of the Higgs boson is uniquely determined at tree level: the
couplings to fermions and gauge bosons are set by their masses and all
production cross sections, decay widths and branching ratios can be
calculated unambiguously \cite{REV}. Unfortunately, $M_{H^0}$ is a
free parameter.  {}From the direct search at LEP1 and LEP2 we know that it
should be larger than about 71 GeV \cite{LEPbound}. Triviality restricts 
the Higgs particle to be lighter than about 1 TeV; theoretical arguments based
on Grand Unification at a scale $\sim10^{16}$ GeV suggest however,
that the preferred mass region will be 100 GeV $ \lsim M_{H^0} \lsim$
200 GeV; for a recent summary, see Ref.~\cite{DATA}.  \s

In supersymmetric (SUSY) theories, the Higgs sector is extended to
contain at least two isodoublets of scalar fields. In the
Minimal Supersymmetric Standard Model (MSSM) this leads to
the existence of five physical Higgs particles: two CP-even Higgs bosons $h$ 
and $H$, one CP-odd or pseudoscalar Higgs boson $A$, and two charged Higgs
particles $H^\pm$ \cite{HHG}. Besides the four masses, two additional
parameters are needed: the ratio of the two vacuum expectation values,
$\tb$, and a mixing angle $\alpha$ in the CP-even sector. However, only
two of these parameters are independent: choosing the pseudoscalar
mass $M_A$ and $\tb$ as inputs, the structure of the MSSM Higgs sector
is entirely determined at lowest order. However, large SUSY radiative
corrections 
\cite{RC} affect the Higgs masses and couplings, introducing new
[soft SUSY-breaking] parameters in the Higgs sector. If in addition relatively
light genuine supersymmetric particles are allowed, the whole set of
SUSY parameters will be needed to describe the MSSM Higgs boson
properties unambiguously. \s 

In this report we describe the program HDECAY\footnote{A complete overview
over all theoretical details can be found in Ref. \cite{habil}.}, which 
calculates the decay widths and branching ratios of Higgs bosons in 
the SM and the MSSM. It includes:

\begin{itemize}

\item All decay channels that are kinematically allowed and which have
  branching ratios larger than $10^{-4}$, {\it y compris} the loop
  mediated, the most important three body decay modes, and in the MSSM
  the cascade and the supersymmetric decay channels.

\item All relevant higher-order QCD corrections to the decays into
  quark pairs and to the quark loop mediated decays into gluons are
  incorporated in the most complete form \cite{QCD}. The largest part
  of the corrections to the heavy quark pair decay modes are mapped
  into running masses which have to be evaluated at the scale of the
  Higgs mass.  The small leading electroweak corrections are also
  included. They become sizeable only in the large Higgs mass regime
  due to the enhanced self-interactions of the Higgs bosons.

\item Double off-shell decays of the CP-even Higgs bosons into massive
  gauge bosons which then decay into four massless fermions
  \cite{2OFF}.  These decays are important for masses close to $M_W$
  or $M_Z$. For larger masses, it is a sufficient approximation to
  switch off these decays [which are CPU time consuming] and to allow
  for one on-shell gauge boson only.

\item All important below-threshold [three-body] decays: with
  off-shell heavy top quarks $H^0, H,A \ra t\bar{t}^* \ra t
  \bar{b}W^-$ and $H^+ \ra t^* \bar b \ra b\bar{b}W^+$; with one
  off-shell gauge boson $H \ra W^{\pm*} H^\mp$, $H\to Z^*A$, $A \ra
  Z^*h$ and $H^\pm \ra W^{\pm*} A, W^{\pm*} h$; as well as the decays
  of $H$ with one off-shell Higgs boson $H \ra hh^*, AA^*$. These
  three body decays can be rather important, especially in the MSSM
  \cite{1OFF} (see also \cite{Moreti}).

\item In the MSSM, the complete radiative corrections in the effective
  potential approach with full mixing in the stop and sbottom sectors;
  it uses the renormalization group improved values of the Higgs
  masses and couplings, and the relevant leading next-to-leading-order
  corrections are also implemented \cite{SUBH}.

\item In the MSSM, all the decays into SUSY particles when they are
  kinematically allowed. The decays into charginos and neutralinos are
  included in the most general case, and the decays to sleptons and
  squark pairs with sfermion mixing in the third generation sector
  \cite{DSUSY}.

\item In the MSSM, all SUSY particles are included in the loop
  mediated $\gamma \gamma$ and $gg$ decay channels: charged Higgs
  bosons, chargino, slepton and squark [including mixing] loops in
  $h,H \ra \gamma \gamma$ decays, chargino loops in $A \ra \gamma
  \gamma$ and squark loops in $h,H \ra gg$. In the gluonic decay modes
  the large QCD corrections for quark \cite{higgsqcd,hgg} and squark
  loops \cite{SQCD} are also included.

\end{itemize}

The basic input parameters, fermion and gauge boson masses and their
total widths, coupling constants and in the MSSM, soft SUSY-breaking
parameters can be chosen from an input file. In this file several
flags allow to switch on/off or change some options [{\it e.g.} choose
a particular Higgs boson, include/exclude the multi-body or SUSY
decays, or include/exclude specific higher-order QCD corrections]. The
results for the many decay branching ratios and the total decay widths
are written into several output files with headers indicating the
processes and giving the input parameters. \s

The program is written in FORTRAN and has been tested on several
machines: VAX stations under the operating system VMS and work
stations running under UNIX. All the necessary subroutines [e.g. for
integration] are included. The program is lengthy [more than 5000
FORTRAN lines] but rather fast, especially if some options [as decays
into double off-shell gauge bosons] are switched off. \s

The rest of this report is organized as follows. In the next section
we discuss the physical decay processes included in the program.  We
describe the parameters of the input file in Section 3. In Section 4,
we present examples of output files. Some comments and conclusions are
given in Section 5.

\section{Decay Modes}

\subsection{Standard Model}

\nn {\bf a) Decays to quarks and leptons} \s

%              ======================================================

The Higgs boson partial width for decays to massless quarks, directly
coupled to the SM Higgs particle, is calculated including the ${\cal
  O}( \alpha_{s}^{3})$ QCD radiative corrections \cite{drees,russ} in
the ${\overline{\rm MS}}$ renormalization scheme. Large logarithms are
resummed by using the running quark mass $\overline{m}_{Q}(M_{H})$ and
the strong coupling constant $\alpha_s(M_H)$ both defined at the scale
of the Higgs boson mass. The quark masses can be neglected in the
phase space and in the matrix element except for decays in the
threshold region, where the next-to-leading-order QCD corrections are given in
terms of the quark {\it pole} mass $M_Q$ \cite{drees}. \s

The relation between the perturbative {\it pole} quark mass ($M_Q$)
and the running $\overline{\rm MS}$ mass (${\overline{m}}_{Q}$) at the
scale of the pole mass can be expressed as \cite{broad}
\begin{equation} \label{run-pole}
{\overline{m}}_{Q}(M_{Q})= \frac{M_{Q}}{\displaystyle 1+\frac{4}{3}
\frac{\alpha_{s}(M_Q)}{\pi} + K_Q \left(\frac{\alpha_s(M_Q)}{\pi}\right)^2}
\end{equation}
where the numerical values of the NNLO coefficients are given by
$K_t\sim 10.9$, $K_b \sim 12.4$ and $K_c \sim 13.4$.  Since the
relation between the pole mass $M_{c}$ of the charm quark and the
${\overline{\rm MS}}$ mass ${\overline{m}}_{c}(M_{c})$ evaluated at
the pole mass is badly convergent \cite{broad}, the running quark
masses ${\overline{m}}_{Q}(M_{Q})$ are adopted as starting points, because
these are directly determined from QCD spectral sum rules \cite{narison} for
the $b$ and $c$ quarks.
The flag NNLO(I) determines whether (I=1) the input running mass is
related to the pole mass according to the eq.~(\ref{run-pole}) or
(I=0) using the simplified version with the $K_Q$ term neglected [in
this case we denote the pole mass by $M_Q^{\rm pt2}$].  The input pole
mass values and corresponding running masses are presented in Table 1.
\begin{table}[hbt]
\renewcommand{\arraystretch}{1.5}
\begin{center}
\begin{tabular}{|c||c|c|c|} \hline
$Q$ & $\overline{m}_Q (M_Q)$ & $M_Q^{\rm pt2}$ & $M_Q$ \\ \hline \hline
$c$ & 1.23 GeV               & 1.41 GeV    & 1.64 GeV    \\
$b$ & 4.23 GeV               & 4.62 GeV    & 4.87 GeV    \\
$t$ & 167.4 GeV              & 175.0 GeV   & 177.1 GeV   \\ \hline
\end{tabular}
\renewcommand{\arraystretch}{1.2}
\caption[]{\it Quark mass values for the $\overline{\rm MS}$ mass and 
  the two different definitions of the pole masses. The strong coupling 
  has been chosen as $\alpha_s(M_Z)=0.118$ and the bottom and charm mass
  values are taken from Ref.~\cite{narison}.}
\end{center}
\end{table}
The evolution from $M_{Q}$ upwards to a renormalization scale $\mu$ is
given by
\begin{eqnarray}
{\overline{m}}_{Q}\,(\mu )&=&{\overline{m}}_{Q}\,(M_{Q})
\,\frac{c\,[\alpha_{s}\,(\mu)/\pi ]}{c\, [\alpha_{s}\,(M_{Q})/\pi ]}
\label{eq:msbarev}
\end{eqnarray}
with \cite{runmass}
\begin{eqnarray*}
c(x)&=&\left(\frac{9}{2}\,x\right)^{\frac{4}{9}} \, [1+0.895x+1.371\,x^{2}]
\hspace{1.35cm} \mbox{for} \hspace{.2cm} M_{s}\,<\mu\,<M_{c}\\
c(x)&=&\left(\frac{25}{6}\,x\right)^{\frac{12}{25}} \, [1+1.014x+1.389\,x^{2}]
\hspace{1.0cm} \mbox{for} \hspace{.2cm} M_{c}\,<\mu\,<M_{b}\\
c(x)&=&\left(\frac{23}{6}\,x\right)^{\frac{12}{23}} \, [1+1.175x+1.501\,x^{2}]
\hspace{1cm} \mbox{for} \hspace{.2cm} M_{b}\,<\mu \,< M_t \\
c(x)&=&\left(\frac{7}{2}\,x\right)^{\frac{4}{7}} \, [1+1.398x+1.793\,x^{2}]
\hspace{1.35cm} \mbox{for} \hspace{.2cm} M_{t}\,<\mu
\end{eqnarray*}
For the charm quark mass the evolution is determined by
eq.~(\ref{eq:msbarev}) up to the scale $\mu=M_b$, while for scales
above the bottom mass the evolution must be restarted at $M_Q = M_b$.
The values of the running $b,c$ masses at the scale $\mu = 100$ \GeV,
characteristic for the Higgs mass, are typically 35\% (60\%) smaller
than the bottom (charm) pole masses $M_b^{\rm pt2}$ ($M_c^{\rm pt2})$.
\s

The program HDECAY includes the full massive NLO corrections close to
the thresholds as well as the massless ${\cal O}(\alpha_s^3)$
corrections far above the thresholds. The transition between both
regions is provided by a linear interpolation as shown in Fig.~1. Thus
the result is optimized for the description of the mass effects in the
threshold region and for the renormalization group improved large
Higgs mass regime. \s
\begin{figure}[hbt]
\vspace*{0.5cm}
\hspace*{-0.5cm}
\begin{turn}{-90}%
\epsfxsize=9cm \epsfbox{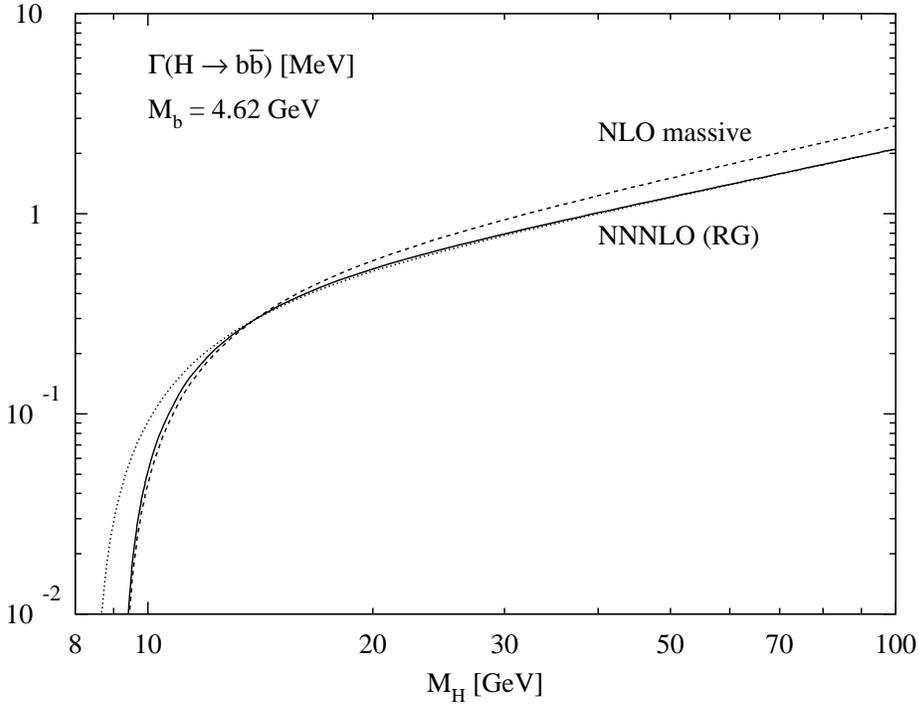}
\end{turn}
\vspace*{-0.2cm}
\caption[ ]{\it Interpolation between the full massive NLO expression (dashed
  line) for the $b\bar b$ decay width of the Standard Higgs boson and
  the renormalization group improved NNNLO result (dotted line). The
  interpolated curve is presented by the full line.}
\end{figure}

The electroweak corrections to heavy quark and lepton decays in the
intermediate Higgs mass range are small \cite{hqqelw} and could thus
be neglected, but the bulk of the effect \cite{hqqelwapp} is included
in the program. For large Higgs masses the electroweak corrections due
to the enhanced self-coupling of the Higgs bosons are included, which
however turn out to be small \cite{hqqlam}.

In the case of $t\bar t$ decays of the Standard Higgs boson, the
${\cal O} (\alpha_s)$ QCD corrections are included according to
\cite{drees}.  Below-threshold (three-body) decays $H\to t\bar t^* \to
t\bar b W^-$ into off-shell top quarks may be sizeable \cite{1OFF} and
thus are implemented. \s

\nn {\bf b) Decays to gluons} \s

%              =======================

The decay of the Higgs boson to gluons is mediated by heavy quark
loops in the Standard Model; the partial decay width in lowest order
is given in \cite{hgglo}.  QCD radiative corrections
\cite{higgsqcd,hgg} are built up by the exchange of virtual gluons,
gluon radiation from the internal quark loop and the splitting of a
gluon into unresolved two gluons or a quark-antiquark pair. The
radiative corrections are very large, nearly doubling the partial
width.  Since $b$ quarks, and eventually $c$ quarks, can in principle
be tagged experimentally, it is physically meaningful to include gluon
splitting $g^*\,\ra\, b{\overline{b}} \; (c{\overline{c}})$ in
$H\ra\,gg^*\ra\, gb{\overline{b}} \, (c{\overline{c}})$ decays to the
inclusive decay probabilities $\Gamma(H \ra b\bar{b}+ \dots)$ {\it
  etc.} \cite{QCD}. Separating this contribution generates large
logarithms, which can be effectively absorbed by defining the number
of active flavors in the gluonic decay mode in the input file of
HDECAY by specifying the NF-GG parameter. The contributions of the
subtracted flavors will automatically be added to the corresponding
heavy quark decay modes. \s

\nn {\bf c) Decays to $\gamma \gamma$ and $Z \gamma$} \s

%              ===========================
The decay of the Higgs boson to two photons and to a photon and a $Z$
boson, mediated by $W$ and heavy fermion loops, are implemented
according to \cite{hgagalo}.  QCD radiative corrections are rather
small \cite{higgsqcd,hgaga} and thus neglected in the program. \s

%\pagebreak[4]

\nn {\bf d) Decays to  $WW/ZZ$ bosons} \s
%           ===============================

Above the $WW$ and $ZZ$ decay thresholds the partial decay widths into
pairs of on-shell massive gauge bosons are given in \cite{hvv}.
Electroweak corrections are small in the intermediate mass range
\cite{hvvelw} and thus neglected in the program HDECAY. Higher order
corrections due to the self-couplings of the Higgs particles are
sizeable \cite{hvvlam} for $M_H \gsim 400$ GeV and are taken into
account. \s

Below the $WW/ZZ$ threshold, the decay modes into off-shell gauge
bosons are important. With the input parameter ON-SH-WZ=1 the program
includes decays with one on-shell and one off-shell gauge boson
\cite{hvvp}, while for ON-SH-WZ=0 decays with both off-shell are also
calculated \cite{2OFF}. The branching ratios for the latter reach the
percent level for Higgs masses above about 100 (110) GeV for both
$W(Z)$ boson pairs off-shell. For higher masses, it is sufficient to
allow for one off-shell gauge boson only, especially because the two
virtual gauge boson option is CPU time consuming.

\subsection{The Minimal Supersymmetric Standard Model}

%           =========================================
The MSSM Higgs sector is implemented in HDECAY including the complete
radiative corrections due to top/bottom quark and squark loops within
the effective potential approach. Next-to-leading order QCD
corrections and the full mixing in the stop and sbottom sectors are
incorporated. The Higgs boson mass spectrum, the mixing angles and
Higgs boson couplings are calculated using the approximate formulae of
M. Carena, M. Quiros and C.E.M. Wagner \cite{SUBH}. The basic
parameters describing the effective Higgs potential at higher orders
are specified in the input file. The formulae for the decay widths at
tree-level have been derived in Ref.~\cite{A1}. \s

\nn {\bf a) Decays to quarks and leptons} \s
%           =============================

The calculation of the partial decay widths of scalar neutral Higgs
bosons $h$ and $H$ to fermions in the MSSM is performed using the same
approximations and options as in the Standard Model case with properly
modified Higgs boson couplings. For massless quarks the QCD
corrections for scalar, pseudoscalar and charged Higgs boson decays
are implemented analogously to the SM case \cite{drees,russ}, $i.e.$ the
Yukawa and QCD couplings are evaluated at the scale of the Higgs boson
mass. \s

In the threshold regions mass effects play a significant role, in
particular for the pseudoscalar Higgs boson, which has an $S$-wave
behavior $\propto \beta$ as compared with the $P$--wave suppression
$\propto \beta^3$ for CP-even Higgs bosons
[$\beta=(1-4m_f^2/M_\Phi^2)^{1/2}$ is the velocity of the decay
fermions].  The QCD corrections to the partial decay width of the
CP-odd Higgs boson $A$ into heavy quark pairs are taken from
Ref.~\cite{drees}, and for the charged Higgs particles from
Ref.~\cite{hud}.  The transition from the threshold region, involving
mass effects, to the renormalization group improved large Higgs mass
regime is provided by a smooth linear interpolation analogous to the
SM case. \s

Below the $t\bar t$ threshold, decays of the neutral Higgs bosons into
off-shell top quarks are sizeable, thus modifying the profile of the
Higgs particles significantly. Off-shell pseudoscalar branching ratios
reach a level of a few percent for masses above about 300 GeV for
small $\tb$ values. Similarly, below the $t\bar b$ threshold,
off-shell decays $H^+\to t^*\bar b \to b\bar b W^+$ are important,
reaching the percent level for charged Higgs boson masses above about
100 GeV for small $\tb$ values. These decays are incorporated
according to the expressions from Ref.~\cite{1OFF}.\s

\nn {\bf b) Decays to gluons} \s
%           ===================

Since the $b$ quark couplings to the Higgs bosons may be strongly
enhanced and the $t$ quark couplings suppressed in the \MSSM, $b$
loops can contribute significantly to the Higgs-$gg$ couplings so that
the approximation $M_{Q}^{2} \gg M_{H}^{2}$ cannot be applied any more
for $M_\Phi \lsim 150$~GeV, where this decay mode is important.
Nevertheless, it turns out {\it a posteriori} that this is an
excellent approximation for the QCD corrections in the range, where these
decay modes are relevant. The LO width for $h,H
\ra gg$ is generated by quark and squark loops with the latter
contributing significantly for Higgs masses below about 400 GeV
\cite{SQCD}. The partial decay widths are calculated according to
Ref.~\cite{higgsqcd,hgg}.  The bottom and charm final states from
gluon splitting can be added to the corresponding $b\bar b$ and $c\bar
c$ decay modes, as in the SM case, by defining NF-GG=3 in the input
file.\s

\nn {\bf c) Decays into $\gamma \gamma$ and $Z\gamma$} \s
%           ========================

The decays of the neutral Higgs bosons to two photons and a photon
plus a $Z$ boson are mediated by $W$ and heavy fermion loops, as in
the Standard Model, and in addition by charged Higgs, sfermion and
chargino loops; the partial decay widths are calculated according to
Ref.~\cite{higgsqcd}. QCD corrections to the quark and squark loop
contributions are small \cite{higgsqcd,hgaga} and thus neglected in
the program.\s

\nn {\bf d) Decays to $WW/ZZ$ gauge bosons} \s
%           =====================================

The partial widths of the CP-even neutral MSSM Higgs bosons into $W$
and $Z$ boson pairs are obtained from the SM Higgs decay widths by
rescaling with the corresponding MSSM couplings. They are strongly
suppressed [due to kinematics in the case of $h$ and reduced couplings
for the heavy $H$], thus not playing a dominant role as in the SM
case. \s

\nn {\bf e) Decays to Higgs boson pairs} \s
%           ===========================

The heavy CP-even Higgs particle can decay into a pair of light
scalars as well to a pair of pseudoscalar Higgs bosons, $H \ra hh$ and
$H \ra AA$.  While the former is the dominant decay mode of $H$ in the
mass range $2M_h < M_H <2m_t$ for small values of $\tb$, the latter
mode occurs only in a marginal area of the MSSM parameter space. For
large values of $\tb$, these decays occur only if $M_A \sim M_h \lsim
M_H/2$, corresponding to the lower end of the heavy Higgs mass range,
and have branching ratios of 50\% each.  Since the $Hb\bar b$ Yukawa
coupling is strongly enhanced for large $\tb$, below threshold decays
$H \ra hh^*, AA^*$ with $A,h \ra b\bar{b}$ are included \cite{1OFF}.
The lightest CP-even Higgs particle $h$ can also decay into
pseudoscalar Higgs pairs for values $\tb \sim 1$ and $M_h <50$ GeV;
however this area of the parameter space is already ruled out by
present data \cite{LEPbound}. \s

\nn {\bf f) Decays to $W/Z$ and Higgs bosons} \s

The Higgs bosons can also decay into a gauge boson and a lighter Higgs
boson.  The branching ratios for the two body decays $A\ra hZ$ and
$H^+ \ra W^+h$ may be sizeable in specific regions of the MSSM
parameter space [small values of $\tb$ and below the $tt/tb$
thresholds for neutral/charged Higgs bosons]. The expressions of the
decay widths are given in e.g.\ Ref.~\cite{1OFF}. \s

Below-threshold decays into a Higgs particle and an off-shell gauge
boson turned out to be rather important for the heavy Higgs bosons of
the MSSM.  Off-shell $A \ra hZ^*$ decays are important for the
pseudoscalar Higgs boson for masses above about 130 GeV for small
$\tb$. The decay modes $H^\pm \to hW^*, AW^*$ reach branching ratios
of several tens of percent and lead to a significant reduction of the
dominant branching ratio into $\tau\nu$ final states to a level of
60\% to 70\% for small $\tb$. In addition, three-body $H \ra AZ^*$ and
$H\ra H^+W^{-*}$, which are kinematically forbidden at the two-body
level, can be sizeable for small $M_A$ values.  The partial decay
widths for these three-body decays are calculated according to the
formulae given in Ref.~\cite{1OFF}.\s

\nn {\bf g) Decays to charginos and neutralinos} \s
%           ==========================

The lightest charginos and neutralinos are expected to have masses of
the order of the $Z$ boson mass. The heavy CP-even, CP-odd and charged
Higgs bosons of the MSSM can therefore decay into these states
\cite{A1}. Present experimental bounds on the SUSY particle masses, do
not allow decays for SUSY decay modes of the lightest CP-even Higgs
boson $h$, except maybe for the decays into a pair of the lightest
neutralinos.  These decays, the partial widths of which can be found
in Ref.~\cite{DSUSY}, are included in the program. \s

The masses of charginos and neutralinos as well as their couplings to
the Higgs bosons depend on three extra parameters [in addition to
those describing the Higgs sector at the tree-level]: the
Higgs-Higgsino mass parameter $\mu$ [which also enters the radiative
corrections in the Higgs sector], the Bino and Wino mass parameters
$M_1$ and $M_2$. Assuming a common gaugino mass at the unification
scale, the parameter $M_1$ is related to $M_2$ by the GUT relation
$M_1=\frac{5}{3}M_2\tan^2\theta_W$.  \s

The chargino and neutralino mass matrices are diagonalized using the
analytical expressions given in Ref.~\cite{A2}. The masses and the
couplings to the Higgs bosons are calculated in the subroutine
GAUGINO. \s

\nn {\bf h) Decays to sleptons and squarks} \s
%           ==========================

The MSSM Higgs bosons can also decay into the SUSY partners of leptons
and quarks if the latter are light enough. In particular, if
kinematically allowed, decays into third generation sfermions can be
dominant due to enhanced couplings \cite{A3}. For instance, the
couplings of the CP-even Higgs bosons to stop pairs are proportional
to $m_t^2/M_Z^2$ and can lead to very large decay widths. \s

The sfermions masses and couplings to Higgs bosons will depend on
three extra parameters [in addition to $\tb$ and $M_A$] for each
generation: the left- and right-handed soft SUSY-breaking mass
parameters $M_{\tilde{f}_L}$ and $M_{\tilde{f}_R}$, the Higgs mass parameter
$\mu$ and the trilinear
coupling $A_f$.  The trilinear couplings are important only in the
case of the third generation sfermions, and only $A_t, A_b$ and
$A_\tau$ need to be introduced. The latter couplings [at least $A_t$
and $A_b$] also contribute to the radiative corrections to the Higgs
sector. For the SUSY breaking scalar masses, we assumed degeneracy in
the first and second generation and treated the third generation
separately\footnote{ We could have taken the same inputs for the three
  generations. However, to allow for a comparison with ISAJET
  \cite{isajet}, we have used different inputs for the SUSY breaking
  scalar mass of the 1st/2nd and the 3rd generation.}. While the
masses of the left- and right-handed 1st/2nd generation sfermions
correspond to the physical sfermion masses, in the third generation
mixing between these fields needs to be included to obtain the
physical eigenstates \cite{A4}. \s

The masses of the sfermions, as well as their couplings to Higgs bosons,
including the mixing in the generation are calculated in the subroutine
SFERMION. The decay widths are calculated in the main subroutine using 
the formulae given in Ref.~\cite{DSUSY}. The QCD corrections to squark
decays [in particular stop and sbottom decays] have been calculated in
Ref.~\cite{sqqcd} but are not yet implemented in the program. 

\section{How to Run HDECAY: Input File}

The HDECAY program is self-contained with all necessary subroutines
included.  In addition to the source code of the program HDECAY, an
input file, defined as unit 98, is needed from which the program reads
the input parameters. The name of this input file can be specified in
the first OPEN statement of HDECAY. It should be noted that the input
numbers must {\it not} start before the equality signs in each
corresponding line. The input file contains the following parameters
[all non-integer parameters are in double precision and the mass
parameters as well as the decay widths and the trilinear couplings are
in GeV]:
\begin{description}
\item \underline{HIGGS:} integer, chooses the Higgs boson to be 
                            considered \\
0: Standard Model Higgs boson $H^0$ \\
1: light CP-even MSSM Higgs boson $h$ \\
2: heavy CP-even MSSM Higgs boson $H$ \\
3: pseudoscalar Higgs boson $A$ \\
4: charged MSSM Higgs bosons $H^\pm$ \\
5: all MSSM Higgs bosons
\item \underline{TGBET:} ratio of the vacuum expectation values in the
  MSSM, $\tb$, the second basic input of the model; the program is
  suitable only for values $\tb \gsim 1$.
\item \underline{MABEG}: start value of the Higgs mass in GeV
\item \underline{MAEND}: end value of the Higgs mass in GeV
\item \underline{NMA}: integer, number of iterations for the input Higgs 
mass\\
In the SM, MA$\equiv M_{H^0}$ while in the MSSM case MA is the pseudoscalar 
Higgs mass $M_A$, which will be the  basic input parameter for the MSSM 
Higgs sector. 
\item \underline{ALS(MZ)}: strong coupling constant at the scale $M_Z$:
$\alpha_S(M_Z)$
\item \underline{MSBAR(1)}: $\overline{\rm MS}$ mass of the strange quark 
at the scale $Q=1$ GeV 
\item \underline{MC}: charm quark pole mass 
\item \underline{MB}: bottom quark pole mass 
\item \underline{MT}: top quark pole mass 
\item \underline{MTAU}: $\tau$ lepton mass 
\item \underline{MMUON}: muon mass 
\item \underline{1/ALPH}: inverse QED coupling constant: $\alpha^{-1}(0)$
\item \underline{GF}: Fermi decay constant 
\item \underline{GAMW}:     total decay width of the $W$ boson
\item \underline{GAMZ}:     total decay width of the $Z$ boson
\item \underline{MZ}:       $Z$ boson mass 
\item \underline{MW}:       $W$ boson mass 
\item \underline{VUS}:      CKM parameter $V_{us}$
\item \underline{VCB}:      CKM parameter $V_{cb}$
\item \underline{VUB/VCB}:  ratio of the CKM parameters 
$V_{ub}/V_{cb}$. 
\item \underline{MU}: SUSY breaking Higgs mass parameter $\mu$
\item \underline{M2}: SUSY breaking gaugino mass parameter $M_2$
\item \underline{MSL1}: SUSY breaking mass parameter for 1st/2nd 
generation left-handed sleptons, $M_{\tilde{l}_L}$
\item \underline{MER1}: SUSY breaking mass parameter for 1st/2nd generation
right-handed charged sleptons, $M_{\tilde{e}_R}$
\item \underline{MSQ1}: SUSY breaking mass parameter for 1st/2nd generation
left-handed up and down type squarks, $M_{\tilde{q}_L}$
\item \underline{MUR1}: SUSY breaking mass parameter for 1st/2nd generation
right-handed up-type squarks, $M_{\tilde{u}_R}$
\item \underline{MDR1}: SUSY breaking mass parameter for 1st/2nd generation
right-handed down-type squarks, $M_{\tilde{d}_R}$
\item \underline{MSL}: SUSY breaking mass parameter for 3rd generation
left-handed sleptons, $M_{\tilde{L}_L}$
\item \underline{MER}: SUSY breaking mass parameter for 3rd generation
right-handed sleptons, $M_{\tilde{\tau}_R}$
\item \underline{MSQ}: SUSY breaking mass parameter for 3rd generation
left-handed up- and down-type squarks, $M_{\tilde{Q}_L}$
\item \underline{MUR}: SUSY breaking mass parameter for right-handed
stops, $M_{\tilde{t}_R}$
\item \underline{MDR}: SUSY breaking mass parameter for right-handed 
sbottoms, $M_{\tilde{b}_R}$
\item \underline{AL}:  SUSY breaking trilinear coupling for $\tau$ 
sleptons, $A_\tau$ 
\item \underline{AU}:  SUSY breaking trilinear coupling for stops, $A_t$
\item \underline{AD}:  SUSY breaking trilinear coupling for sbottoms, $A_b$
\item \underline{NNLO (M)}: integer, \\
=0: use ${\cal O}(\alpha_s)$ formula for the quark pole masses $\ra
\overline{\rm MS}$ masses\\
=1: use ${\cal O}(\alpha_s^2)$ formula for the quark pole masses $\ra
\overline{\rm MS}$ masses
\item \underline{ON-SHELL}: integer \\
=0: include three-body decays with off-shell tops, Higgs and gauge 
bosons\\
=1: exclude three-body decays with off-shell tops, Higgs and gauge bosons
\item \underline{ON-SH-WZ}: integer \\
=0: include double off-shell decays to gauge bosons $\Phi \ra W^*W^*,Z^*Z^*$ 
\\ 
=1: include only single off-shell gauge bosons $\Phi \ra W^*W,Z^*Z $ 
\item \underline{IPOLE}: integer \\
=0: calculate $\overline{\rm MS}$ masses of the MSSM Higgs particles \\
=1: calculate pole masses of the MSSM Higgs particles
\item \underline{OFF-SUSY}: integer \\
=0: include decays into and loops of supersymmetric particles \\
=1: exclude decays into and loops of supersymmetric particles
\item \underline{INDIDEC}: integer \\
=0: write out only the sums of chargino, neutralino and sfermion decays \\
=1: write out all individual chargino, neutralino and sfermion decays
\item \underline{NF-GG:} integer \\
number of light flavors included in the decays
$\Phi \ra gg^* \ra g q \bar{q}$ (NF-GG=3,4 or 5).
\end{description}

The current values of the SM parameters [fermion masses, gauge boson
masses and total widths, coupling constants, CKM angles] are given in
Tab.~2, where an example of the input file is displayed. The entire
Higgs sector of the MSSM is fixed once the parameters $\tb, M_A , \mu
, M_2$, the masses $M_{\tilde{L}_L}, M_{\tilde{E}_R}, M_{\tilde{U}_L},
M_{\tilde{U}_R}, M_{\tilde{D}_R}$ and the trilinear couplings $A_\tau
, A_t$ and $A_b$ are specified. Some examples for these values are
shown in Tab.~2.
\begin{table}[hbtp]
\vspace*{-0.3cm}
\begin{verbatim}
HIGGS    = 0
TGBET    = 1.5D0
MABEG    = 100.D0
MAEND    = 500.D0
NMA      = 5
ALS(MZ)  = 0.118D0
MSBAR(1) = 0.190D0
MC       = 1.42D0
MB       = 4.62D0
MT       = 175.D0
MTAU     = 1.7771D0
MMUON    = 0.105658389D0
1/ALPHA  = 137.0359895D0
GF       = 1.16639D-5
GAMW     = 2.080D0
GAMZ     = 2.490D0
MZ       = 91.187D0
MW       = 80.33D0
VUS      = 0.2205D0
VCB      = 0.04D0
VUB/VCB  = 0.08D0
MU       = 300.D0
M2       = 200.D0
MSL1     = 500.D0
MER1     = 500.D0
MQL1     = 500.D0
MUR1     = 500.D0
MDR1     = 500.D0
MSL      = 500.D0
MER      = 500.D0
MSQ      = 500.D0
MUR      = 500.D0
MDR      = 500.D0
AL       = 1500.D0
AU       = 1500.D0
AD       = 1500.D0
NNLO (M) = 0
ON-SHELL = 0
ON-SH-WZ = 0
IPOLE    = 0
OFF-SUSY = 1
INDIDEC  = 0
NF-GG    = 5
\end{verbatim}
\vspace*{-0.88cm}
\nn \centerline{Table~2: \it Example of the input file.}
\end{table}

\section{Results of Test Run: Output Files} 

The output is written to several files. Only the output files of the
chosen HIGGS boson(s) are printed, and they contain all decay
branching ratios and the total decay width, except for the decays to
SUSY particles [if OFF-SUSY=0] where only the sums of the branching
ratios into charginos, neutralinos, sleptons and squarks are printed,
if the flag INDIDEC=0; only for INDIDEC=1 all individual branching
ratios are printed in additional output files. For convenience, an
output file br.input is printed in which the input parameters are
given.  Below we will describe the output files in the SM and the MSSM
[also with the option for SUSY decays switched on] and 
list all the decay channels which we have
considered for the various Higgs bosons.

\subsection{Standard Model Higgs boson}

For the SM Higgs boson, in addition to the file br.input for the input
parameters, two output files are printed in which the total decay
width and the following 11 branching ratios are given [notice that we 
have put the decays into fermions and gauge bosons into two different 
files] 
\beq
{\rm br.sm1}:  && M_{H^0} \ , \ BR(b\bar{b})   \ , \ BR(\tau^+ \tau^-) 
\ , \ BR(\mu^+ \mu^-) \ , \ BR(s\bar{s})   \ , \ BR(c\bar{c})  \ , \
BR(t\bar{t}) \non \\
{\rm br.sm2}:  && M_{H^0} \ , \ BR(gg)   \ , \ BR(\gamma \gamma) 
\ , \ BR(\gamma Z) \ , \ BR(WW)   \ , \ BR(ZZ)  \ , \ \Gamma_{H^0}^{\rm 
tot} \non 
\eeq
For the example of input file shown in Tab.~2, one
obtains the two outputs given in Tab.~3. The various branching ratios
and the total decay width are shown in Fig.~2.
\begin{table}[hbtp]
\begin{small}
\begin{verbatim}
   MHSM        BB       TAU TAU     MU MU         SS         CC         TT
___________________________________________________________________________

 100.000     0.8119     0.7926E-01 0.2752E-03 0.6048E-03 0.3698E-01     0.
 200.000     0.2596E-02 0.2884E-03 0.1000E-05 0.1928E-05 0.1177E-03     0.
 300.000     0.6082E-03 0.7274E-04 0.2521E-06 0.4513E-06 0.2754E-04 0.5293E-04
 400.000     0.2283E-03 0.2869E-04 0.9940E-07 0.1694E-06 0.1033E-04 0.1376
 500.000     0.1183E-03 0.1542E-04 0.5342E-07 0.8772E-07 0.5347E-05 0.1936

   MHSM          GG     GAM GAM     Z GAM         WW         ZZ       WIDTH
___________________________________________________________________________

 100.000     0.5807E-01 0.1532E-02 0.4654E-04 0.1025E-01 0.1046E-02 0.2598E-02
 200.000     0.8219E-03 0.5241E-04 0.1753E-03 0.7350     0.2609      1.428
 300.000     0.5674E-03 0.1289E-04 0.5670E-04 0.6913     0.3073      8.510
 400.000     0.7532E-03 0.3192E-05 0.1935E-04 0.5872     0.2741      28.89
 500.000     0.5476E-03 0.4897E-06 0.7666E-05 0.5450     0.2607      67.53
\end{verbatim}
\end{small}
\centerline{Table~3: \it The two output files in the SM with the 
inputs of Tab.~2.}
\end{table}

\begin{figure}[hbtp]
\begin{center}
\vspace*{-1.0cm}

\hspace*{-1.5cm}
\begin{turn}{-90}%
\epsfxsize=9.cm \epsfbox{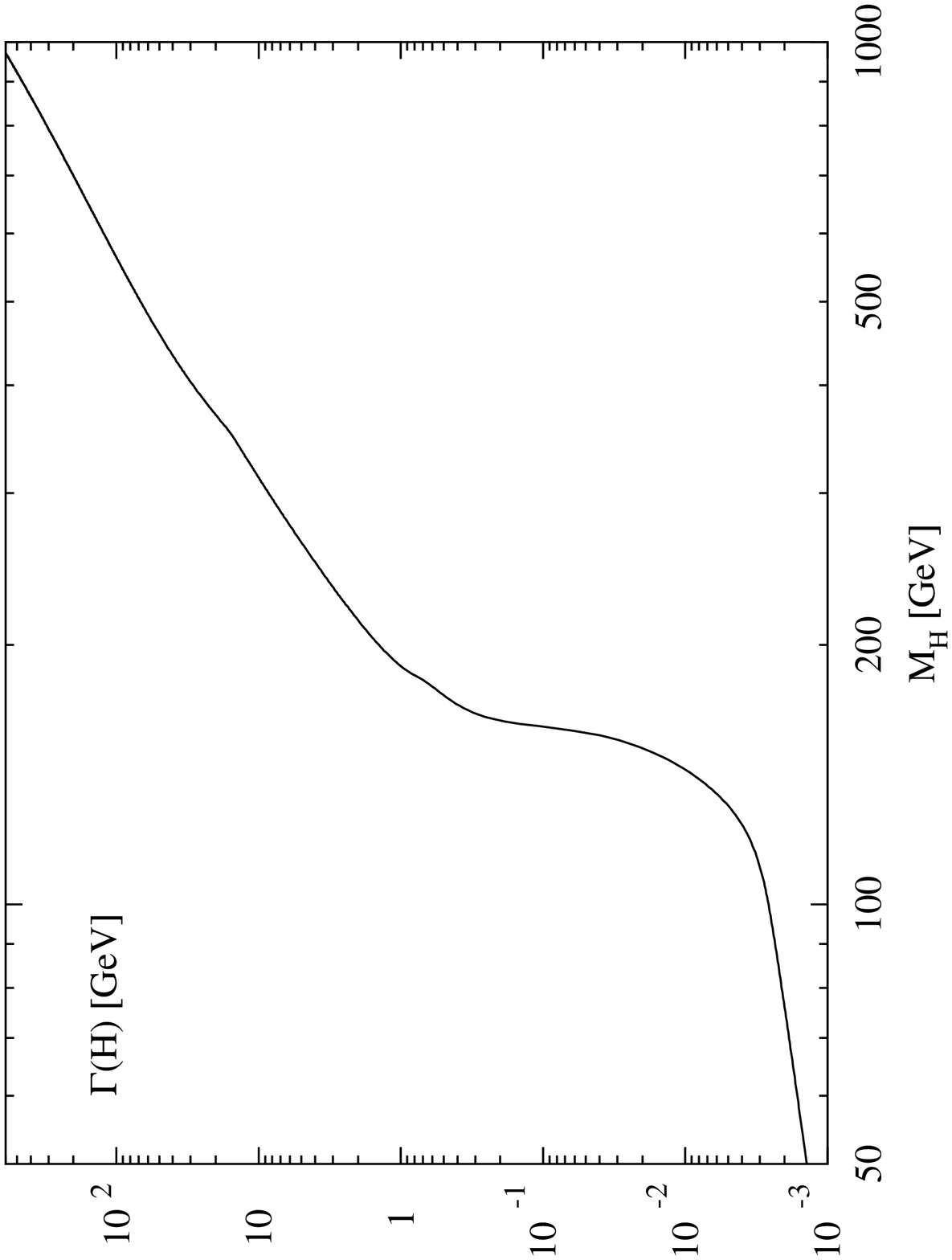}
\end{turn}
\vspace*{1.cm}

\hspace*{-1.5cm}
\begin{turn}{-90}%
\epsfxsize=9.cm \epsfbox{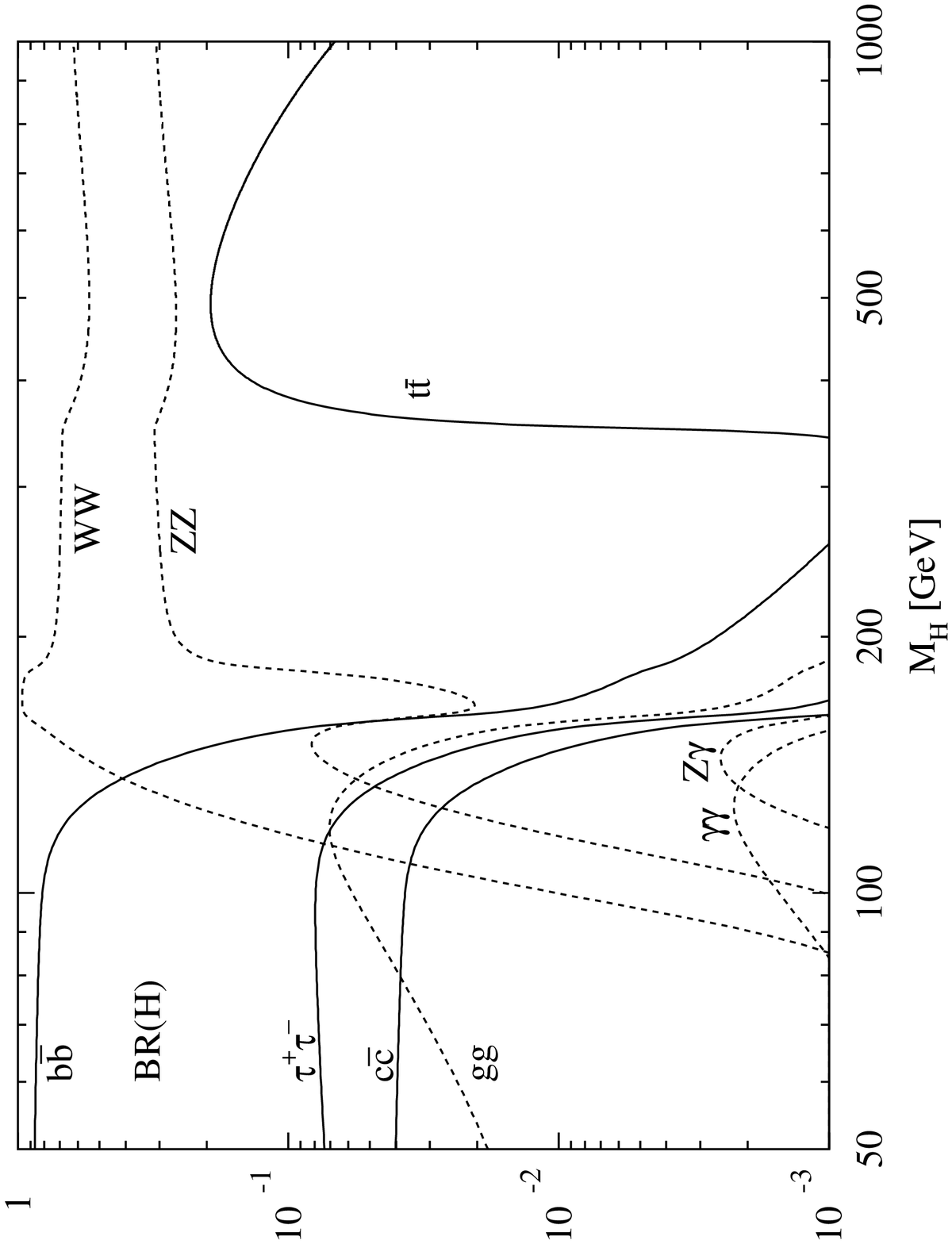}
\end{turn}
\vspace*{0.0cm}
\end{center}
\caption[]{\it Total decay width $\Gamma(H)$ in GeV and the main branching
ratios $BR(H)$ of the Standard Model Higgs decay channels, using the inputs
of Tab.~2.}
\end{figure}

%\newpage

\subsection{MSSM without SUSY decays}

As discussed earlier, the two basic inputs of the program for the MSSM
Higgs sector are $\tb$ and $M_A$. Once these parameters are fixed, all
the other Higgs masses and couplings are determined at tree-level.
However, subleading effects due to squark mixing [mainly the parameters
$A_{t,b}$ and $\mu$] will alter these values.

\bigskip

\underline{For the lightest MSSM Higgs boson $h$}, the decays are 
the same as in the SM if the SUSY decays are switched off. The
two output files contain the following branching ratios
\beq
{\rm br.l1}:  && M_h \ , \ BR(b\bar{b})   \ , \ BR(\tau^+ \tau^-) 
\ , \ BR(\mu^+ \mu^-) \ , \ BR(s\bar{s})   \ , \ BR(c\bar{c})  \ , \
BR(t\bar{t}) \non \\
{\rm br.l2}:  && M_h \ , \ BR(gg)   \ , \ BR(\gamma \gamma) 
\ , \ BR(\gamma Z) \ , \ BR(WW)   \ , \ BR(ZZ)  \ , \ \Gamma_h^{\rm tot} \non 
\eeq

\bigskip

\underline{For the heavy CP-even MSSM Higgs boson $H$}, there are more
possibilities for the decays due to the larger mass: in addition to
the same decay modes as $h$, cascade decays into two light Higgs
bosons or a mixed pair of Higgs and gauge bosons occur. The branching
ratios of 15 decay modes are written to the following output files
\beq
{\rm br.h1}:  && M_H \ , \ BR(b\bar{b})   \ , \ BR(\tau^+ \tau^-) 
\ , \ BR(\mu^+ \mu^-) \ , \ BR(s\bar{s})   \ , \ BR(c\bar{c})  \ , \
BR(t\bar{t}) \non \\
{\rm br.h2}:  && M_H \ , \ BR(gg)   \ , \ BR(\gamma \gamma) 
\ , \ BR(\gamma Z) \ , \ BR(WW)   \ , \ BR(ZZ)  \non \\
{\rm br.h3}:  && M_H \ , \ BR(hh)   \ , \ BR(AA) 
\ , \ BR(AZ) \ , \ BR(W^\pm H^\mp)   \ , \ \Gamma_H^{\rm tot} \non 
\eeq

\bigskip

\underline{For the CP-odd MSSM Higgs boson $A$}, there are less
possibilities than for the $H$ boson: due to CP-invariance, the
pseudoscalar $A$ does not couple to gauge and Higgs boson pairs.
The 10 decay channels are printed in the output files as follows
\beq
{\rm br.a1}:  && M_A \ , \ BR(b\bar{b})   \ , \ BR(\tau^+ \tau^-) 
\ , \ BR(\mu^+ \mu^-) \ , \ BR(s\bar{s})   \ , \ BR(c\bar{c})  \ , \
BR(t\bar{t}) \non \\
{\rm br.a2}:  && M_A \ , \ BR(gg)   \ , \ BR(\gamma \gamma) 
\ , \ BR(\gamma Z) \ , \ BR(hZ)   \ , \ \Gamma_A^{\rm tot} \non 
\eeq

\bigskip

\underline{For the charged MSSM Higgs bosons $H^\pm$}, there are
8 decay channels which can exceed the $10^{-4}$ level; these are written 
in the two output files
\beq
{\rm br.c1}:  && M_{H^+} \ , \ BR(c\bar{b})   \ , \ BR(\tau^+ \nu_\tau) 
\ , \ BR(\mu^+ \nu_\mu ) \ , \ BR(u\bar{s})   \ , \ BR(c\bar{s})  \ , \
BR(t\bar{b}) \non \\
{\rm br.c2}:  && M_{H^+} \ , \ BR(hW^+)   \ , \ BR(A W^+) \ , \ 
\Gamma_{H^+}^{\rm tot} \non 
\eeq

\smallskip

\begin{table}[hbtp]
\begin{small}
\begin{verbatim}
   MHL         BB       TAU TAU     MU MU         SS         CC         TT
__________________________________________________________________________
 70.7080     0.9058     0.8281E-01 0.2938E-03 0.6587E-03 0.5552E-02     0.
 85.8094     0.8728     0.8282E-01 0.2935E-03 0.6372E-03 0.2018E-01     0.
 89.2134     0.8531     0.8150E-01 0.2888E-03 0.6242E-03 0.2827E-01     0.
 90.4005     0.8438     0.8079E-01 0.2863E-03 0.6180E-03 0.3196E-01     0.
 90.9499     0.8390     0.8041E-01 0.2849E

   MHL           GG     GAM GAM     Z GAM         WW         ZZ       WIDTH
___________________________________________________________________________
 70.7080     0.4520E-02 0.2694E-03     0.     0.7392E-04 0.2137E-04 0.4399E-02
 85.8094     0.2164E-01 0.7488E-03     0.     0.7356E-03 0.1789E-03 0.3241E-02
 89.2134     0.3353E-01 0.9854E-03     0.     0.1421E-02 0.3015E-03 0.2778E-02
 90.4005     0.3927E-01 0.1090E-02     0.     0.1833E-02 0.3638E-03 0.2607E-02
 90.9499     0.4226E-01 0.1143E-02     0.     0.2072E-02 0.3972E-03 0.2528E-02
\end{verbatim}
\end{small}

\nn {Table~4a: \it The output files br.l1 and br.l2 for the light CP-even 
$h$ decays without SUSY particles using the inputs in Tab.~2 but with 
HIGGS=5.}

\begin{small}
\begin{verbatim}
   MHH         BB       TAU TAU     MU MU         SS         CC         TT
__________________________________________________________________________
 145.680     0.3669     0.3930E-01 0.1390E-03 0.2575E-03 0.2475E-01     0.
 221.034     0.5570E-01 0.6430E-02 0.2274E-04 0.3953E-04 0.9850E-03     0.
 313.454     0.7133E-01 0.8781E-02 0.3105E-04 0.5072E-04 0.8778E-03 0.7003E-02
 409.924     0.5451E-02 0.7035E-03 0.2487E-05 0.3878E-05 0.5864E-04 0.9421
 507.876     0.2584E-02 0.3460E-03 0.1223E-05 0.1839E-05 0.2610E-04 0.9784

   MHH           GG     GAM GAM     Z GAM         WW         ZZ
__________________________________________________________________________
 145.680     0.1001     0.3088E-03 0.9334E-03 0.2607     0.3433E-01
 221.034     0.1002E-01 0.1167E-04 0.5215E-04 0.3527     0.1420
 313.454     0.2296E-01 0.5107E-04 0.2259E-04 0.2488     0.1116
 409.924     0.4751E-02 0.1359E-04 0.3069E-05 0.1211E-01 0.5670E-02
 507.876     0.2782E-02 0.8875E-05 0.1970E-05 0.3967E-02 0.1900E-02

   MHH           hh         AA        Z A     W+- H-+     WIDTH
___________________________________________________________________________
 145.680     0.1508     0.7363E-06 0.2011E-01 0.1428E-02 0.5795E-02
 221.034     0.4320     0.2673E-10 0.3105E-04 0.4103E-06 0.1237
 313.454     0.5285     0.2548E-12 0.2991E-05 0.1732E-07 0.1496
 409.924     0.2910E-01 0.6195E-15 0.3915E-07 0.1561E-09  2.564
 507.876     0.9974E-02 0.2049E-16 0.4845E-08 0.1594E-10  6.600
\end{verbatim}
\end{small}

\nn {Table~4b: \it The output files br.h1, br.h2 and br.h3 for the 
heavy CP-even $H$ decays without SUSY particles using the inputs in 
Tab.~2 but with HIGGS=5.} 
\end{table}
\begin{table}[hbtp]
\begin{small}
\begin{verbatim}
   MHA         BB       TAU TAU     MU MU         SS         CC         TT
__________________________________________________________________________
 100.000     0.8790     0.8610E-01 0.3046E-03 0.6401E-03 0.7925E-02     0.
 200.000     0.4748     0.5313E-01 0.1879E-03 0.3449E-03 0.4270E-02     0.
 300.000     0.2582     0.3118E-01 0.1102E-03 0.1874E-03 0.2321E-02 0.1953E-01
 400.000     0.2185E-02 0.2777E-03 0.9816E-06 0.1585E-05 0.1964E-04 0.9892
 500.000     0.1714E-02 0.2263E-03 0.8001E-06 0.1244E-05 0.1541E-04 0.9926

   MHA           GG     GAM GAM     Z GAM       Z HL      WIDTH
__________________________________________________________________________
 100.000     0.2539E-01 0.5914E-04 0.8400E-07 0.5965E-03 0.5413E-02
 200.000     0.7001E-01 0.1944E-03 0.2781E-04 0.3970     0.1755E-01
 300.000     0.1403     0.4287E-03 0.9264E-04 0.5476     0.4487E-01
 400.000     0.4515E-02 0.1468E-04 0.3641E-05 0.3820E-02  6.718
 500.000     0.3162E-02 0.1090E-04 0.2810E-05 0.2275E-02  10.30
\end{verbatim}
\end{small}

\nn {Table~4c: \it The output files br.a1 and br.a2 for the CP-odd $A$ 
decays without SUSY particles using the inputs in Tab.~2 but with HIGGS=5.}

\bigskip

\begin{small}
\begin{verbatim}
   MHC         BC       TAU NU      MU NU         SU         CS         TB
__________________________________________________________________________
 126.847     0.9519E-02 0.6288     0.2224E-02 0.2059E-03 0.5637E-01 0.5893E-01
 214.686     0.2091E-04 0.1521E-02 0.5378E-05 0.4520E-06 0.1237E-03 0.9654
 309.984     0.4888E-05 0.3800E-03 0.1343E-05 0.1057E-06 0.2892E-04 0.9914
 407.542     0.3382E-05 0.2755E-03 0.9740E-06 0.7309E-07 0.2001E-04 0.9955
 506.054     0.2899E-05 0.2448E-03 0.8654E-06 0.6265E-07 0.1715E-04 0.9971

   MHC           hW         hA      WIDTH
__________________________________________________________________________
 126.847     0.2142     0.2965E-01 0.9404E-03
 214.686     0.3296E-01 0.2509E-05 0.6581
 309.984     0.8185E-02 0.6596E-07  3.804
 407.542     0.4160E-02 0.9100E-08  6.898
 506.054     0.2599E-02 0.2187E-08  9.640
\end{verbatim}
\end{small}

\nn {Table~4d: \it The output files br.c1 and br.c2 for the charged  $H^+$ 
decays without SUSY particles using the inputs in Tab.~2 but with HIGGS=5.}
\end{table}
Examples of the output files for the four MSSM Higgs bosons excluding the 
SUSY decays are shown in Tab.~4a--d. The numbers are obtained by using the 
input file of Tab.~2, with  HIGGS=5. 

\begin{figure}[hbtp]

\vspace*{-2.5cm}
\hspace*{-4.5cm}
\begin{turn}{-90}%
\epsfxsize=16cm \epsfbox{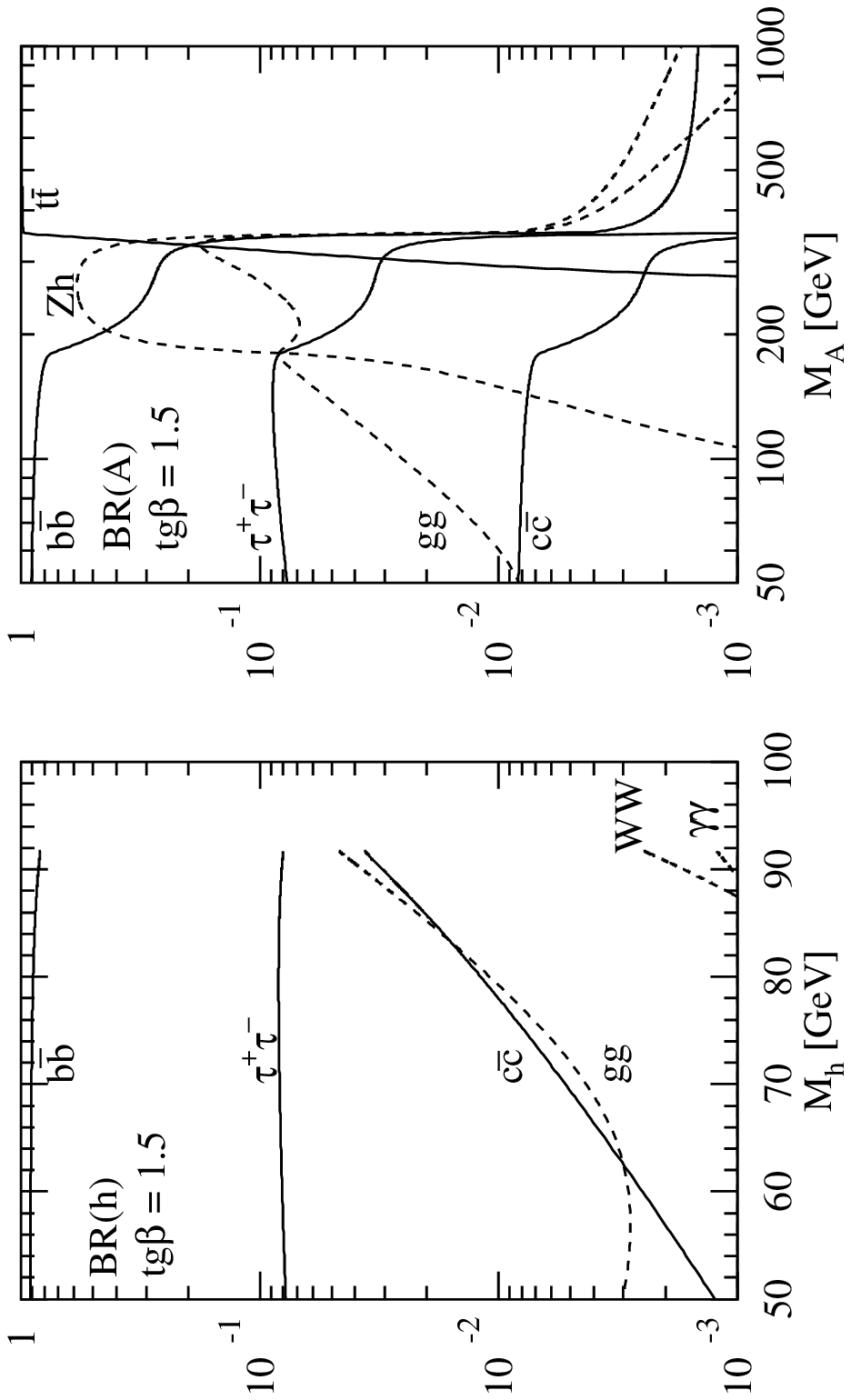}
\end{turn}
\vspace*{-4.2cm}

\centerline{\bf Fig.~\ref{fg:mssmbr}a}

\vspace*{-2.5cm}
\hspace*{-4.5cm}
\begin{turn}{-90}%
\epsfxsize=16cm \epsfbox{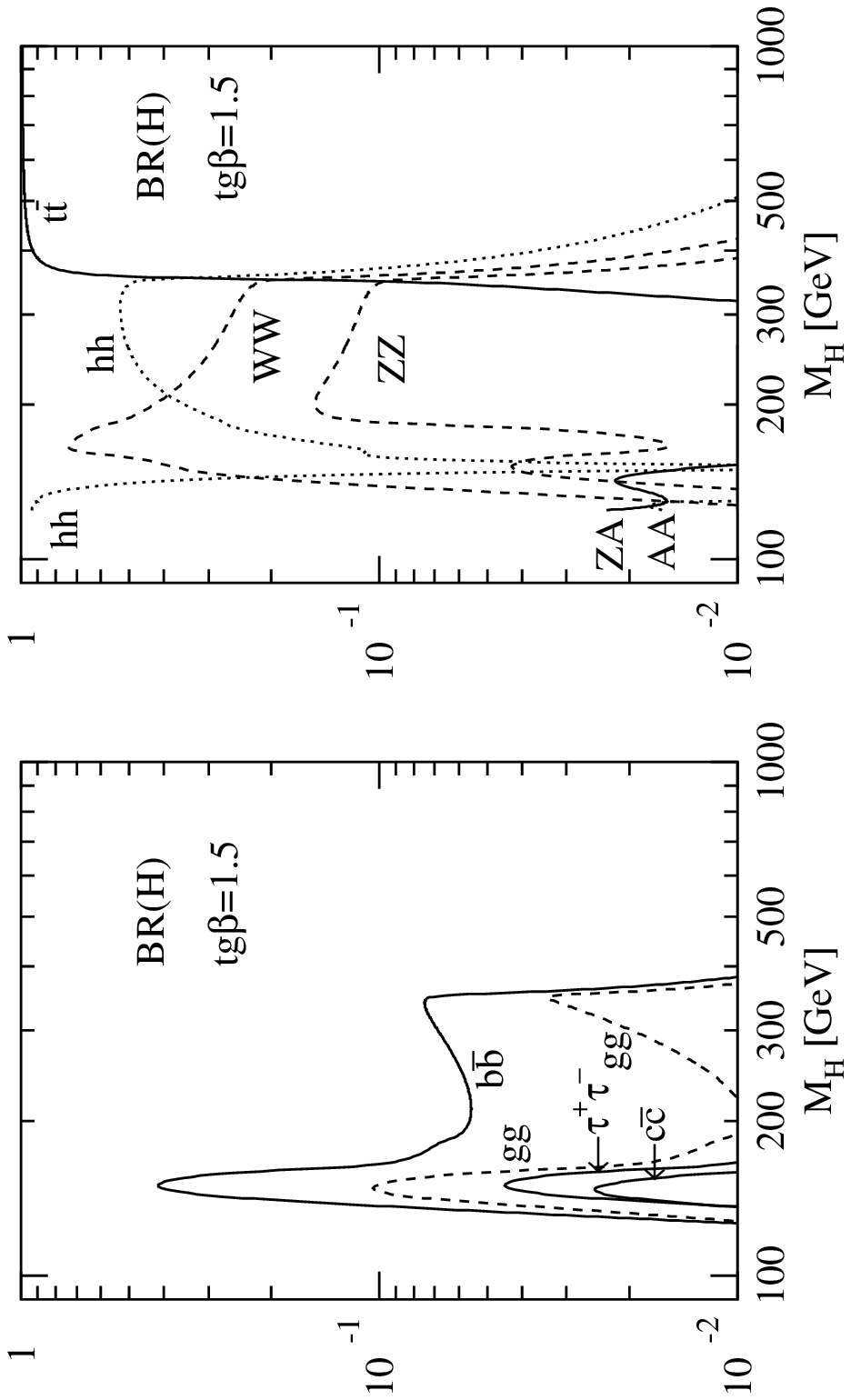}
\end{turn}
\vspace*{-4.2cm}

\centerline{\bf Fig.~\ref{fg:mssmbr}b}

\caption[]{\label{fg:mssmbr} \it Branching ratios of the MSSM Higgs bosons
$h, A (a), H (b), H^\pm (c)$ and their total decay widths $\Gamma(\Phi) (c)$,
using the inputs of Tab.~2.}
\end{figure}
\addtocounter{figure}{-1}
\begin{figure}[hbtp]

\vspace*{-2.5cm}
\hspace*{-4.5cm}
\begin{turn}{-90}%
\epsfxsize=16cm \epsfbox{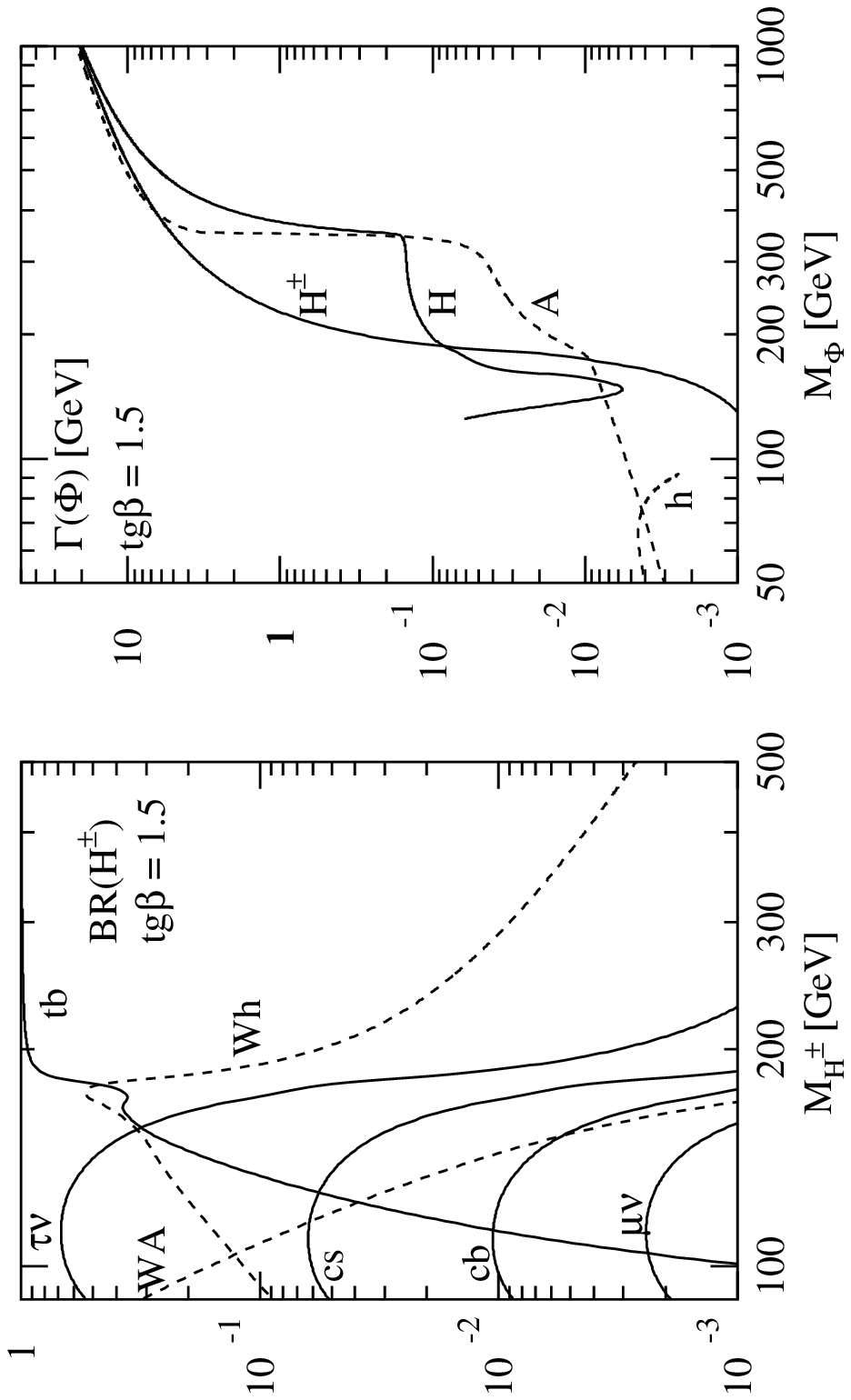}
\end{turn}
\vspace*{-4.2cm}

\centerline{\bf Fig.~\ref{fg:mssmbr}c}

\caption[]{\it Continued.}
\end{figure}
The branching ratios for the main decay channels [those listed in the output
files] of $h, H,A$ and $H^+$ as function of corresponding masses 
are shown in Fig.~3a--c.
The total decay widths $\Gamma(\Phi)$ of the four MSSM Higgs bosons are shown
in Fig.~3c.

\subsection{MSSM with SUSY decays}

If the SUSY particles charginos, neutralinos, sleptons and squarks are
relatively light, a plethora of new decay channels is available
especially for the heavy CP-even, CP-odd and charged Higgs bosons. For
the light CP-even Higgs boson $h$, due to the present experimental
bounds on the SUSY particle masses \cite{A5}, the only decays which
could be allowed are decays into the lightest neutralinos, and
possibly decays into the lightest charginos, sleptons and stop squarks
if $M_h$ is close to its maximum allowed value [for large $\tb$ and
$A_t$ values] and the sparticles masses are close to their
experimental bounds. \s

The pseudoscalar Higgs boson decays into all kinds of charginos and
neutralinos, but due to CP-invariance it does not decay into the first
and second generation squarks since we neglect their mixing.
Therefore we have included only the decays into top and bottom squarks
and $\tau$ sleptons $A \ra \tilde{t}_1 \tilde{t}_2, \tilde{b}_1
\tilde{b}_2$ and $\tilde{\tau}_1 \tilde{\tau}_2$. The heavy CP-even
Higgs boson decays into all possible chargino, neutralino, slepton and
squark states. Also, the charged Higgs boson will decay into
chargino-neutralino pairs, slepton and squark pairs; however, the
decays into sfermions will be important only for the third generation.
\s

Because the number of possibilities for decays into SUSY particles is
huge we include in the output only the sum of decays into all
charginos, neutralinos, sleptons and squarks
\beq
{\rm br.Xs}:  && M_{\Phi} \ , \ 
\sum BR(\chi^+_j \chi^-_j) \ , \ 
\sum BR(\chi^0_j \chi^0_j) \ , \ 
\sum BR(\tilde{l}_i \tilde{l}_j ) \  , \ 
\sum BR(\tilde{q}_i \tilde{q}_j ) \non 
\eeq
with X = l,h,a and c for $\Phi=h,H,A$ and $H^\pm$ respectively
[for sleptons and squarks of the two first generations the combinations 
$i \neq j$ do not contribute]. However, the main program calls all 
individual decays, except for the two first generations of squarks
and sleptons for which the decays have been summed. For the flag INDIDEC=1
all the branching ratios of the individual channels are written to the output 
files br.Xsi with $i=1,\ldots,5$. \s

Examples of output files for the heavy CP-even, CP-odd and charged
Higgs boson decays into SUSY particles is shown in Tab.~5a--c [no SUSY
decay channel is allowed for the lightest CP-even Higgs boson].  The
numbers are obtained by using the input file of Tab.~2, with HIGGS=5
and OFF-SUSY=0. For convenience, the input SUSY parameters and the
resulting masses for the various SUSY particles are also printed.  One
has to check, that with the choice of input parameters the current
experimental lower bounds on the SUSY particle masses are satisfied
[we do not include any experimental constraint since these bounds will
probably vary in the near future].

\begin{table}[hbtp]
\vspace*{-0.3cm}
\begin{small}
\begin{verbatim}
TB= 1.50000     M2= 200.000     MU= 300.000     MSQ= 500.000
C1=157.197 C2= 343.795 N1= 85.367 N2=162.619 N3= 300.696 N4= 348.902
MST1= 244.202     MST2= 703.413     MSUL= 498.877     MSUR= 499.522
MSB1= 497.816     MSB2= 503.780     MSDL= 501.358     MSDR= 500.239
TAU1= 498.934 TAU2= 502.663 EL= 500.882 ER= 500.716 NL= 498.398
   MHH        CHARGINOS  NEUTRALS   SLEPTONS   SQUARKS
____________________________________________________________________
 145.680         0.         0.         0.         0.
 221.034         0.     0.2379E-02     0.         0.
 313.454         0.     0.3335E-01     0.         0.
 409.924     0.6221E-02 0.6506E-01     0.         0.
 507.876     0.2204E-01 0.3325E-01     0.     0.7003
\end{verbatim}
\end{small}
\vspace*{-0.2cm}

\nn {Table~5a: \it The output file br.hs for the heavy CP-even $H$ 
decays into SUSY particles with the inputs in Tab.~2 with HIGGS=5,
OFF-SUSY=0.}

\begin{small}
\begin{verbatim}
TB= 1.50000     M2= 200.000     MU= 300.000     MSQ= 500.000
C1=157.197 C2= 343.795 N1= 85.367 N2=162.619 N3= 300.696 N4= 348.902
MST1= 244.202     MST2= 703.413     MSUL= 498.877     MSUR= 499.522
MSB1= 497.816     MSB2= 503.780     MSDL= 501.358     MSDR= 500.239
TAU1= 498.934 TAU2= 502.663 EL= 500.882 ER= 500.716 NL= 498.398
   MHA        CHARGINOS  NEUTRALS   SLEPTONS   SQUARKS
_______________________________________________________________________________
 100.000         0.         0.         0.         0.
 200.000         0.     0.5923         0.         0.
 300.000         0.     0.8161         0.         0.
 400.000     0.9247E-01 0.6735E-01     0.         0.
 500.000     0.9382E-01 0.7468E-01     0.         0.
\end{verbatim}
\end{small}
\vspace*{-0.2cm}

\nn {Table~5b: \it The output file br.as for the heavy CP-odd $A$ 
decays into SUSY particles with the inputs in Tab.~2 with HIGGS=5,
OFF-SUSY=0.} 
\vspace*{-0.2cm}

\bigskip

\begin{small}
\begin{verbatim}
TB= 1.50000     M2= 200.000     MU= 300.000     MSQ= 500.000
C1=157.197 C2= 343.795 N1= 85.367 N2=162.619 N3= 300.696 N4= 348.902
MST1= 244.202     MST2= 703.413     MSUL= 498.877     MSUR= 499.522
MSB1= 497.816     MSB2= 503.780     MSDL= 501.358     MSDR= 500.239
TAU1= 498.934 TAU2= 502.663 EL= 500.882 ER= 500.716 NL= 498.398
   MHC        CHARG/NEU  SLEPTONS   SQUARKS
_______________________________________________________________________________
 126.847         0.         0.         0.
 214.686         0.         0.         0.
 309.984     0.2895E-01     0.         0.
 407.542     0.3083E-01     0.         0.
 506.054     0.8753E-01     0.         0.
\end{verbatim}
\end{small}
\vspace*{-0.2cm}

\nn {Table~5c: \it The output file br.cs for the heavy charged $H^+$ 
decays into SUSY particles with the inputs in Tab.~2 with HIGGS=5,
OFF-SUSY=0.} 
\end{table}

\bigskip

Note finally, the numbers included in the standard decay files for the 
$gg$, $\gamma \gamma$ and $Z \gamma$ decay channels will now include
(OFF-SUSY=0) the contribution of the SUSY loops. Furthermore, the total 
width which is quoted at the end includes of course the partial decay 
widths into SUSY particles. 

%\bigskip

\section{Summary and Outlook} 

We have described the Fortran code HDECAY which calculates the total
widths and the branching fractions of the decays of the Higgs bosons
in the Standard Model as well as in its minimal supersymmetric
extension.  In the SM, all decay modes are included; the QCD
corrections to the hadronic decays as well as the possibility of
virtual intermediate states have been incorporated according to the
present state of the art.  In the MSSM, the complete radiative
corrections in the Higgs sector have been implemented in the effective
potential approach. The QCD corrections to the hadronic decays, the
main three-body decay channels as well as the decays into charginos,
neutralinos, squarks and sleptons and the SUSY contributions to the
standard decay modes have been implemented. \s

The program, although lengthy, is fast, and has been tested on several 
machines; it can be easily used. The basic SM and MSSM input parameters 
can be chosen from an input file which contains several flags to switch 
on/off some options as e.g.\ multi-body decays, SUSY particle decays or 
higher-order radiative corrections. Examples for the output files for the 
decay branching ratios with some options have been given. \s

While for the case of the SM the program is rather complete, in the case
of the MSSM the present version of the program can be extended/improved 
in several aspects: 

\begin{itemize}

\item In the present version, the total widths and branching ratios
  are calculated as a function of the Higgs boson mass with other
  parameters kept fixed. In the MSSM, it would be useful to vary other
  parameters such as $\mu, M_2$, {\it etc.}.

\item The QCD corrections to the decays of the heavy MSSM Higgs bosons 
into squark pairs have been calculated recently and found to be rather
large \cite{sqqcd}. These corrections are not yet implemented in the program.

\item There are some three-body decays for heavy Higgs bosons which
can reach the per cent level and which are not yet included. Examples of
such decays are $H^0 \ra WWZ, t\bar{t}Z$ and $t\bar{b}W$ in the SM 
\cite{A6} and $H,A \ra t\bar{t}Z$ and $t\bar{b}W$ in the MSSM. 

\item We have restricted ourselves to the scenario where the lightest 
neutralino is the lightest SUSY particle. Models where the LSP is the
gravitino have been discussed and in this case the Higgs decays into
gravitinos can be very important \cite{gravitino} and should be included. 

\end{itemize}

These extensions and improvements will be made in the next version of the 
program. 
 
\vspace*{1cm}

\nn {\bf Acknowledgments} \s

We would like to thank G.\ Polesello and E.\ Richter-Was for testing
the program on several machines and many useful discussions and
comments.  Thanks also go to Peter Zerwas for suggesting to write this
program.  The work of JK has been partially supported by the Committee
for Scientific Research (Poland) under grant No.\ 2 P03B 180 09.

\newpage

% **************************  References ****************************

\end{document}